\pdfoutput=1
\documentclass[aps,prapplied,superscriptaddress,reprint]{revtex4-2}
\usepackage{graphicx}
\usepackage{dcolumn}
\usepackage{bm}
\usepackage{epstopdf}
\usepackage{color}
\usepackage{hyperref}
\usepackage{amsmath} 
\draft 
\begin{document}
\title{Mid-infrared dispersion relations in InP based photonic crystal slabs revealed by Fourier-transform angle-resolved reflection spectroscopy} 
%
%
%
\author{Siti~Chalimah}
\affiliation{National Institute for Materials Science (NIMS), 1-1 Namiki, Tsukuba 305-0044, Japan}
\affiliation{Graduate School of Engineering, Kyushu University, NIMS, 1-1 Namiki, Tsukuba 305-0044, Japan}
\author{Yuanzhao~Yao}
\author{Naoki~Ikeda}
\affiliation{National Institute for Materials Science (NIMS), 1-1 Namiki, Tsukuba 305-0044, Japan}
\author{Kei~Kaneko}
\author{Rei~Hashimoto}
\author{Tsutomu~Kakuno}
\author{Shinji~Saito}
\affiliation{Corporate Manufacturing Engineering Center, Toshiba Corporation, Isogo, Yokohama 235-0017, Japan}
%
%
\author{Takashi~Kuroda}
\email[Author to whom correspondence should be addressed: ]{kuroda.takashi@nims.go.jp}
\affiliation{National Institute for Materials Science (NIMS), 1-1 Namiki, Tsukuba 305-0044, Japan}
\affiliation{Graduate School of Engineering, Kyushu University, NIMS, 1-1 Namiki, Tsukuba 305-0044, Japan}
\author{Yoshimasa~Sugimoto}
\author{Kazuaki~Sakoda}
\affiliation{National Institute for Materials Science (NIMS), 1-1 Namiki, Tsukuba 305-0044, Japan}

%


\date{\today}

\begin{abstract}
Photonic crystals (PCs) offer unique ways to control light-matter interactions. The measurement of dispersion relations is a fundamental prerequisite if we are to create novel functionalities in PC devices. Angle-resolved spectroscopic techniques are commonly used for characterizing PCs that work in the visible and near-infrared regions. However, the techniques cannot be applied to the mid- and long-wavelength infrared regions due to the limited sensitivity of infrared detectors. Here, we propose an alternative approach to measuring infrared dispersion relations. We construct a high-precision angle-resolved setup compatible with a Fourier-transform spectrometer with an angle resolution as high as 0.3 degrees. Hence, the reflection spectra are mapped to the 2D photonic band structures of In(Ga,Al)As/InP based PC slabs, which are designed as mid-infrared PC surface-emitting lasers. We identify complex PC modes with the aid of polarization selection rules derived by the group theory. Spectral analysis makes it possible to evaluate the mode quality (Q) factors. Therefore, angle-resolved reflection is a useful way of optimizing 2D PC parameters for mid-infrared devices. 
\end{abstract}

\maketitle 

\textcolor{blue}{%
This is the version of the article accepted for publication in \textit{Physical Review Applied}. The final published version will be available from the journal's site as an open access paper. %
}%
\section{Introduction}
Photonic crystal (PC) surface-emitting lasers (PCSELs) are an emerging laser architecture whose unique characteristics include a high output power that scales with the device area, good mode quality, and low beam divergence thanks to the high spatial coherence across the entire device \cite{Imada_APL99,Hirose_NatPhoton2014,Noda_review2017,Ishizaki_review2019}. The principle of the laser action relies on the zero group velocity of light at photonic band edges, which results in strong in-plane feedback inside a PC slab. Moreover, when band edges are formed at the $\varGamma$ point in momentum space, radiation into free space occurs purely out-of-plane, i.e., normal to the slab, as a consequence of momentum conservation rules. Hence, careful PC design that realizes perfect resonance between the band edge mode and material gain frequencies is essential for achieving vertical emission. The measurement of in-plane dispersion relations is thus frequently required to verify optimum conditions in fabricated devices. 

A technique commonly used to characterize photonic band structures in PCSEL devices is the angle-resolved observation of sub-threshold luminescence spectra \cite{Sakai_IEEE05,Williams_PTL12,Taylor_JPhysD2013}. Light waves, which are associated with spontaneous emission from gain media and then coupled to in-plane PC modes, are diffracted into free space along particular directions, which conserve wavevector projections to the slab plane \cite{Astranov_IEE98}. Hence, the angular profiles of the emission spectra are mapped to two-dimensional (2D) photonic band structures, similar to the electronic band structures revealed by angle-resolved photoelectron spectra (ARPES). The technique is widely used for the characterization of visible and near-infrared wavelengths PCSELs. However, the technique cannot be applied to mid-infrared wavelengths due to the limited sensitivity of infrared detectors. To develop mid-infrared PCSELs, which take advantages of novel quantum cascade laser (QCL) technologies \cite{Colombelli_Science03,Boyle_APL16,Liang_APL19,Wang_OEx19}, an alternative scheme is required for characterizing PC slabs. 

Recently, we reported the use of angle-resolved reflection measurement to determine infrared dispersion relations in model PC slabs based on silicon-on-insulator (SOI) waveguides \cite{Yao_OEX20}. We put considerable effort into achieving high angle resolution and developed a high precision variable-angle reflection apparatus that is compatible with a Fourier transform spectrometer \cite{kuroda_arXiv2020}. Here, we apply the angle-resolved reflection technique to the characterization of 
actual PCSEL devices formed of In(Ga,Al)As/InP based QCL multi-layer structures. We identify complex PC modes in the vicinity of the $\varGamma$ point with the aid of rigorous polarization selection rules 
derived by the group theory \cite{sakoda_book}. 
Moreover, spectral analysis makes it possible to evaluate the quality (\textit{Q}) factors of waveguide modes semi-quantitatively. Thus, the angle-resolved reflection measurement is a useful tool with which to determine fundamental parameters for laser actions. 

\section{Experimental procedure}
\subsection{Samples}
Mid-infrared QCL multi-layer structures were grown on $n^{+}$~InP(100) by molecular beam epitaxy. We grew a lattice-matched InGaAs buffer layer with a thickness of 1.0~$\mu$m, an InP bottom cladding layer with a thickness of 2.5~$\mu$m, and a thin InGaAs guiding layer with a thickness of 0.3~$\mu$m, followed by InGaAs/InAlAs strain-compensated multiple quantum wells (MQWs) with a total thickness of 1.6~$\mu$m. The MQWs serve as a QCL active layer that 
has an optical gain at a center wavelength of 4.387~$\mu$m. We then grew a 1.0~$\mu$m thick InGaAs layer, which we used for PC processing. See, the layer sequence and the expected refractive indices in Fig.~\ref{fig_fieldDistributionVertical}(a) in Appendix. 

Finite-element calculations were carried out to determine the target PC design, that is a 2D square lattice of InGaAs cylindrical pillars with a height of 0.8~$\mu$m, a radius of 0.565~$\mu$m, and a lattice constant (\textit{a}) of 1.36~$\mu$m. This design results in a resonance between the measured QCL frequency and the second lowest band edge at the $\varGamma$ point ($\varGamma^{(2)}$) in the first Brillouin zone. (The $\varGamma^{(2)}$ band edge originated from zone folding at the $X$ symmetric point.) 
Here, we deal mainly with the TM-like polarization modes, since a QCL employs inter-subband electronic transitions, which have dipole moments perpendicular to the MQW layer. 

Three millimeter square PC slabs with $a=$ 1.33, 1.36, and 1.39~$\mu$m were fabricated using electron beam lithography and reactive ion beam etching techniques. (SiO$_2$ was used as a mask for the dry etching.) Note that the final PCSEL devices have another InP layer on top, which serves as a top cladding layer and then InGaAs pillars are embedded, followed by a metal gate. We study the samples before InP regrowth and metal gate deposition. 
The laser action of the final PCSEL devices will be reported in a forthcoming article. 

\subsection{Measurement setups}
Mid-infrared angle-resolved reflection measurements were performed using a Fourier transform spectrometer (Jasco, FT/IR-6800) attached to a home-built variable angle apparatus, where infrared light was  carefully collimated, and beam divergence was limited to less than 0.3$^\circ$. The collimated beam passed through an infrared beam splitter, which then transmitted the beam from the sample to the detector. The setup enabled us to observe reflection spectra even at normal incidence, and to change the incident angle ($\theta$) from $-3.7^\circ$ to $+3.7^\circ$ across zero. A wire grid polarizer was inserted into the beam path, and it allowed to resolve \textit{p}- and \textit{s}-polarized components, which have the electric field parallel and perpendicular to the incident plane, respectively. The unique design, which is further detailed in Ref.~\onlinecite{kuroda_arXiv2020}, makes it possible to determine 
the mode frequencies as a function of in-plane wavevector and dispersion relations in the vicinity of the $\varGamma$ point. 

\begin{figure}
\includegraphics[width=8.7cm]{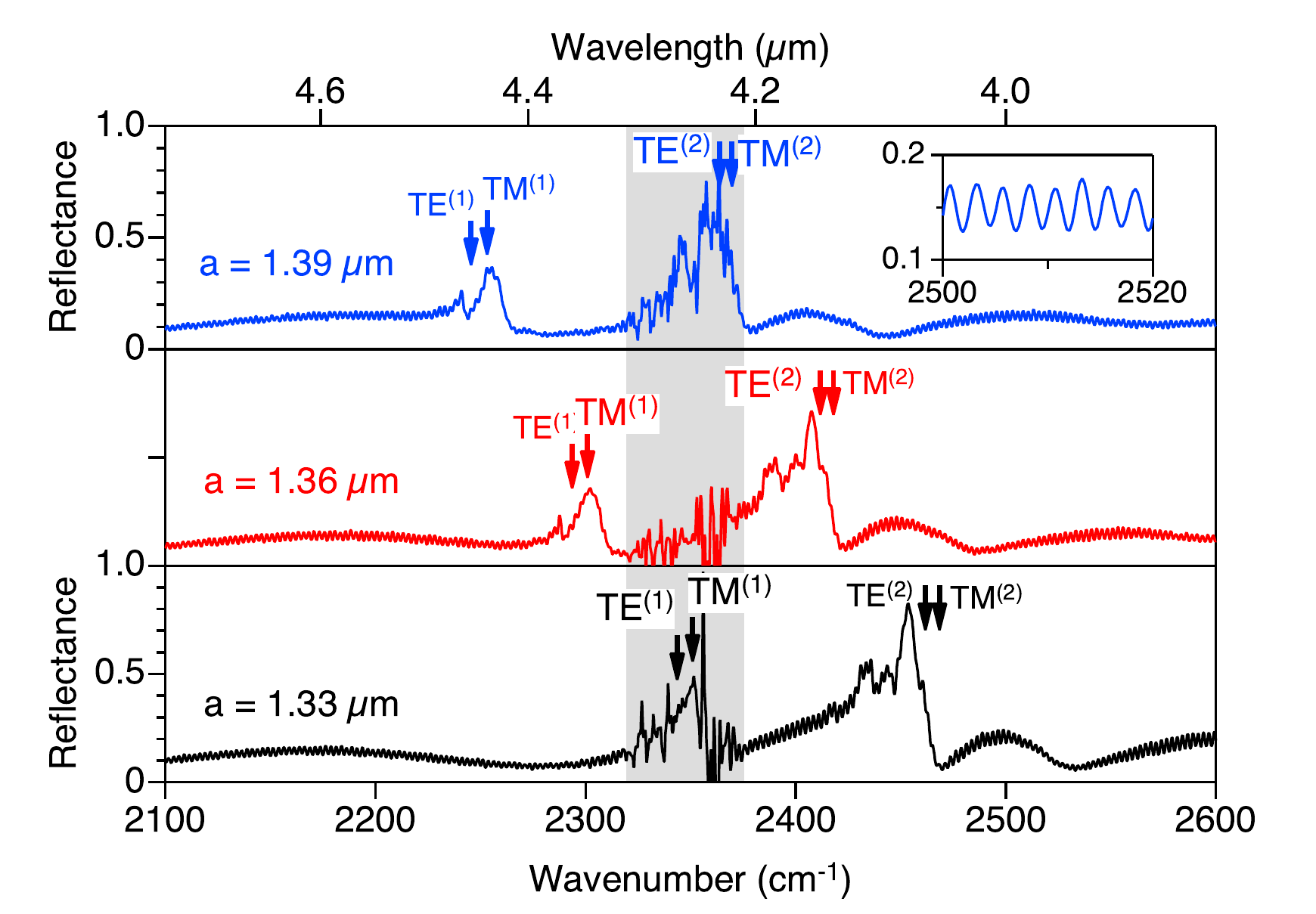}%
\caption{\label{fig_spctrNormal}Normal incidence reflection spectra for samples with different lattice constants, $a=$1.33, 1.36, and 1.39~$\mu$m. The gray region indicates a CO$_2$ absorption band, which causes clear spectral noises. The small sinusoidal modulation superimposed on the spectra is due to Fabry-Perot interference between the top and bottom surfaces of the sample, as highlighted by the expanded spectrum in the inset. }
\end{figure}

\section{Results and discussion}
\subsection{Mid-infrared characterization of PC slabs}
Figure~\ref{fig_spctrNormal} shows reflection spectra at normal incidence ($\theta = 0$) for samples with different lattice constants. All the samples exhibit similar spectral curves, which shift to higher wavenumbers (shorter wavelengths) in unison for smaller lattice constants. (The onset of noisy spectra at 2320-2375~cm$^{-1}$, as shown by the gray region, is due to light absorption by CO$_2$.) The observed scaling behavior is evidence that signature spectra are induced by resonant coupling to PC modes. 

The vertical arrows seen above each spectrum indicate the mode energies calculated by the finite element method. The first and second arrows from the left (low wavenumber side) are the lowest energy TE-like and TM-like modes, respectively. They are denoted by TE$^{(1)}$ and TM$^{(1)}$. The third and fourth arrows are higher order confinement modes (TE$^{(2)}$ and TM$^{(2)}$), which have a nodal field distribution in a direction normal to the slab. (See 
Fig.~\ref{fig_fieldDistributionVertical} in Appendix 
for the field distributions.) These calculated modes are in fairly good agreement with the measured spectral peaks, although the peak intensities are strongly dependent on the mode index, as discussed later. 
A discrepancy between the calculated and the measured mode energies is attributed to the use of inaccurate refractive indices for the constituent materials, particularly that of the bottom InGaAs buffer layer, which contains relatively dense dopants and serves as an antiguiding plasmonic layer. 

\begin{figure}
\includegraphics[width=8.8cm]{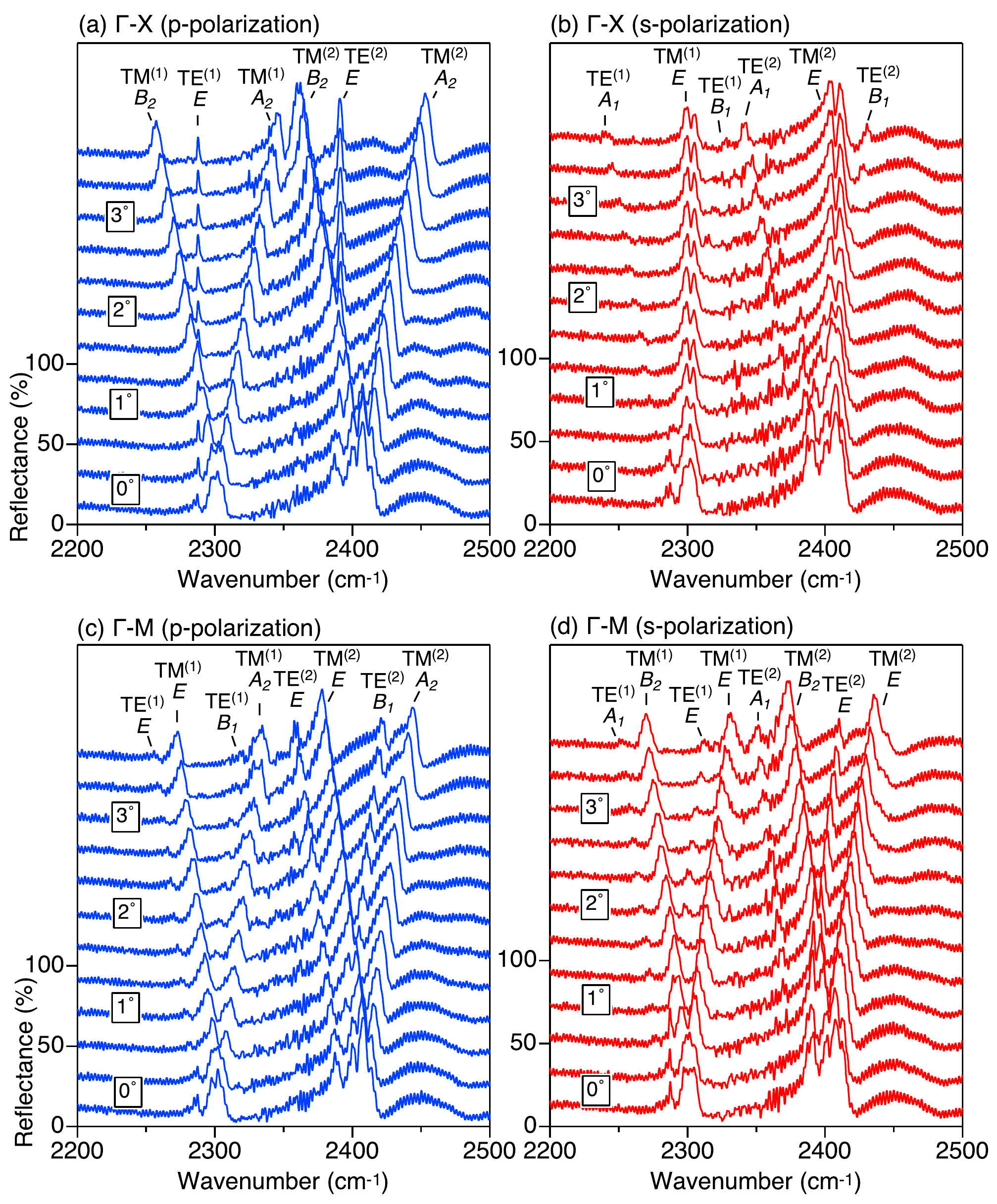}%
\caption{\label{fig_spctrAngle}Angle-resolved reflection spectra for the sample where $a=1.36$~$\mu$m. The spectra in each panel are arranged from bottom to top, $\theta =$~0 to $\sim$3.7$^\circ$ in 0.35$^\circ$ steps. (a) and (b) show \textit{p}- and \textit{s}-polarized spectra, respectively, where the incident beam is tilted inside a plane that contains the [100] axis of the square lattice. Accordingly, the wavevector moves along the $\varGamma$-$X$ axis in the first Brillouin zone. (c) and (d) show the polarized spectra when the wavevector moves along the $\varGamma$-$M$ axis. The peak assignments are also indicated at the top in each panel. %
}
\end{figure}

Figure~\ref{fig_spctrAngle} shows a series of reflection spectra for various incident angles. Here, we study a sample where $a=1.36$~$\mu$m and analyze the spectra for both \textit{s}- and \textit{p} polarizations. In Figs.~\ref{fig_spctrAngle}(a) and \ref{fig_spctrAngle}(b), from bottom to top, the incident beam is tilted from the surface normal towards the [100] in-plane axis, where the wavevector is moved from $\varGamma$ to $X$ in momentum space. 
For \textit{p}-polarization (Fig.~\ref{fig_spctrAngle}(a)), two intense peaks, which were classified as TM$^{(1)}$ and TM$^{(2)}$ at $\theta=0$, are split into two peaks that move to the opposite side. Other small peaks classified as TE$^{(1)}$ and TE$^{(2)}$ stay almost constant. 
In contrast, for \textit{s}-polarization (Fig.~\ref{fig_spctrAngle}(b)) the intense peaks do not move, but small split peaks, which originate from TE$^{(1)}$ and TE$^{(2)}$, are apparent at $\theta \gg 0$. 

In Figs.~\ref{fig_spctrAngle}(c) and \ref{fig_spctrAngle}(d), the beam is tilted towards the [110] axis (45$^\circ$ away from [100]), and the wavevector is moved from $\varGamma$ to $M$. In this geometry, all the peaks split and shift as the angle $\theta$ increases. Polarization dependence is not evident in these plots, but we will analyze the significant energy shift between the \textit{s}- and \textit{p}-polarized spectra. See also the spectral comparison in Fig.~\ref{fig_Qfactor}(a). %



\begin{table}
\caption{\label{tbl_slctn} Selections rules for the observation of spectral peaks around the $\varGamma^{(2)}$ band edge in $C_{4v}$ symmetric PC slabs.}
%
\begin{tabular}{l p{12mm} c p{12mm} c} \hline\hline
Mode\footnotemark[1] & & $\varGamma$-$X$ & & $\varGamma$-$M$ \\ \hline
              $E$    & & $s,\, p$        & & $s,\, p$        \\
              $A_1$  & & $s$             & & $s$             \\
              $A_2$  & & $p$             & & $p$             \\
              $B_1$  & & $s$             & & $p$             \\
              $B_2$  & & $p$             & & $s$             \\ \hline\hline
\end{tabular}%
%
\footnotetext[1]{
These mode symmetries are defined with the symmetry operation of magnetic fields. }%
\end{table}

\subsection{Mode assignment and dispersion relations}
We assign these reflection peaks in terms of irreducible representations of the $C_{4v}$ point group \cite{sakoda_book}. For a uniform slab with no periodic index modulation, the band edge at the $\varGamma^{(2)}$ point has a four-fold degeneracy, 
which is then lifted into a doubly degenerate $E$ mode and two other non-degenerate modes when the $C_{4v}$ symmetric modulation is introduced into the slab. The non-degenerate modes are assigned to $A_1$ and $B_1$ for the TE-like modes, and $A_2$ and $B_2$ for the TM-like modes, where we consider the symmetry operation of magnetic fields. These mode symmetries are consistent with the prediction by the group theory based on the zone folding of TE and TM planar waveguide bands \cite{sakoda_book}. %
Of these modes, only the $E$ mode is optically active at normal incidence, but 
the other modes are inactive due to symmetry mismatching. At off-normal incidence, all the modes become active and split. 
The group theory also predicts polarization selection rules for the emergence of reflection peaks, which are described in Table~\ref{tbl_slctn} \cite{Yao_OEX20_2}. Using this table, we can safely identify all the measured peaks. 
The assignment results can be seen at the top of each panel in Fig.~\ref{fig_spctrAngle}. It is noteworthy that the above-mentioned symmetries are defined at the $\varGamma$ point, and the measured spectral peaks at finite incident angles do not necessarily follow the relevant symmetry characteristics. 


\begin{figure}
\includegraphics[width=8.8cm]{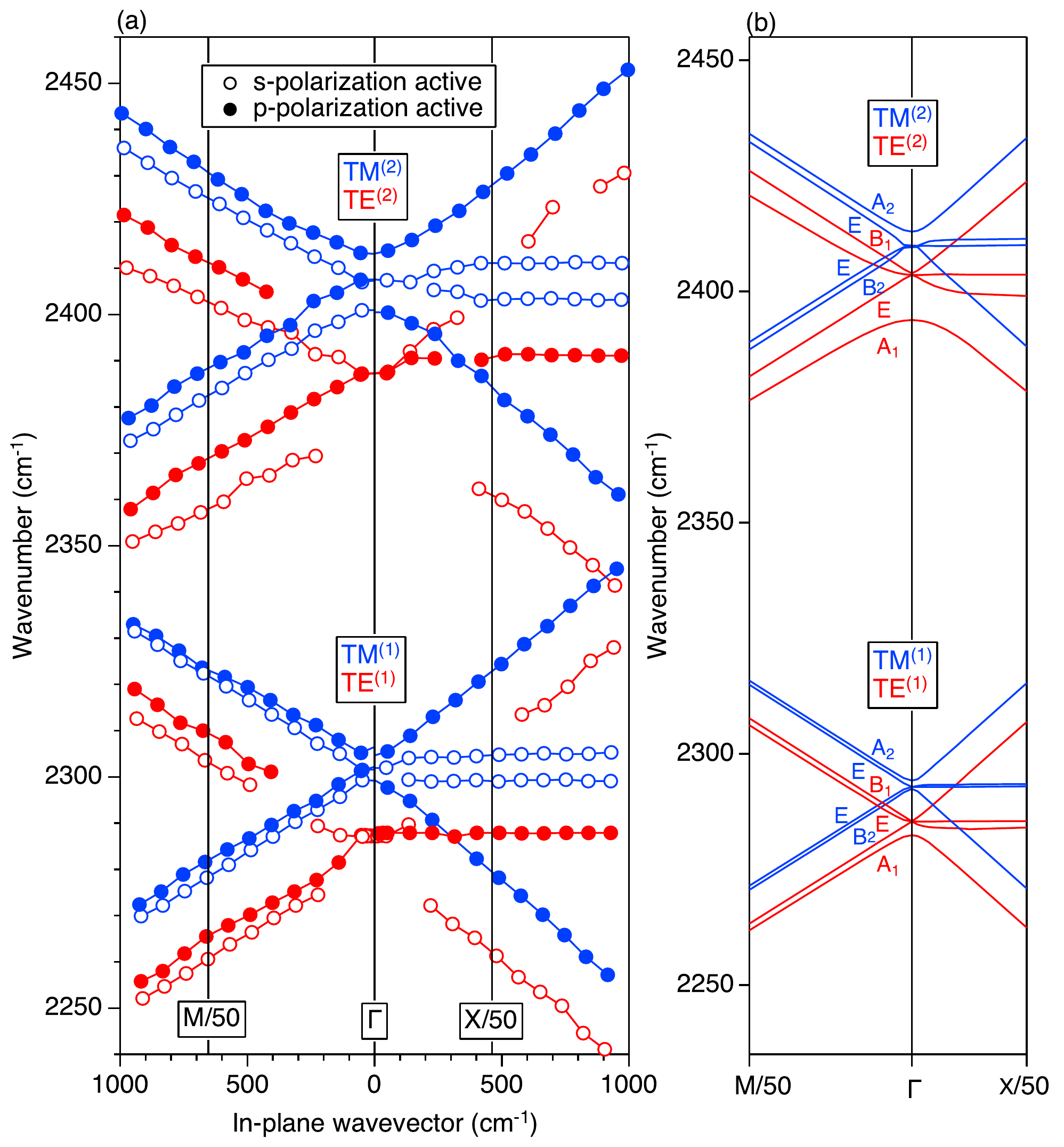}%
\caption{\label{fig_dispersion}(a) Dispersion relations based on measured angle-resolved reflection spectra in Fig.~\ref{fig_spctrAngle}. 
(b) Calculated 2D dispersion relations, where we assume that $a=1.365$~$\mu$m. The abscissa is normalized to $1/50$ of the wavevector at the $X$ band edge ($\pi/50a$) and the $M$ band edge ($\sqrt{2}\pi/50a$).}
\end{figure}

We evaluate the center energies of reflection peaks via fitting. For simplicity we use a multiple Gaussian function as a fitting model. The extracted energies for different $\theta$ values are summarized in the band dispersion curves in Fig.~\ref{fig_dispersion}(a), where we plot the spectral peak energies (in wavenumber units) as a function of a wavevector projection to the slab plane, i.e., $(2\pi/\lambda)\sin\theta$, where $\lambda$ is the vacuum wavelength. For comparison, Fig.~\ref{fig_dispersion}(b) shows calculated band dispersion curves over $1/50$ of the first Brillouin zone. Both measured and calculated curves agree well. Moreover, we analyzed the field distribution of each mode, and confirmed that their symmetry agrees with our mode assignment based on the selection rules. See, Figs.~\ref{fig_fieldDistributionLateral_TE} and \ref{fig_fieldDistributionLateral_TM} and the related discussion in Appendix. Thus, angle-resolved reflection measurement is a useful technique for experimentally determining photonic band structures. 


\subsection{\textit{Q} factor quantification}
\begin{figure}
\includegraphics[width=8.7cm]{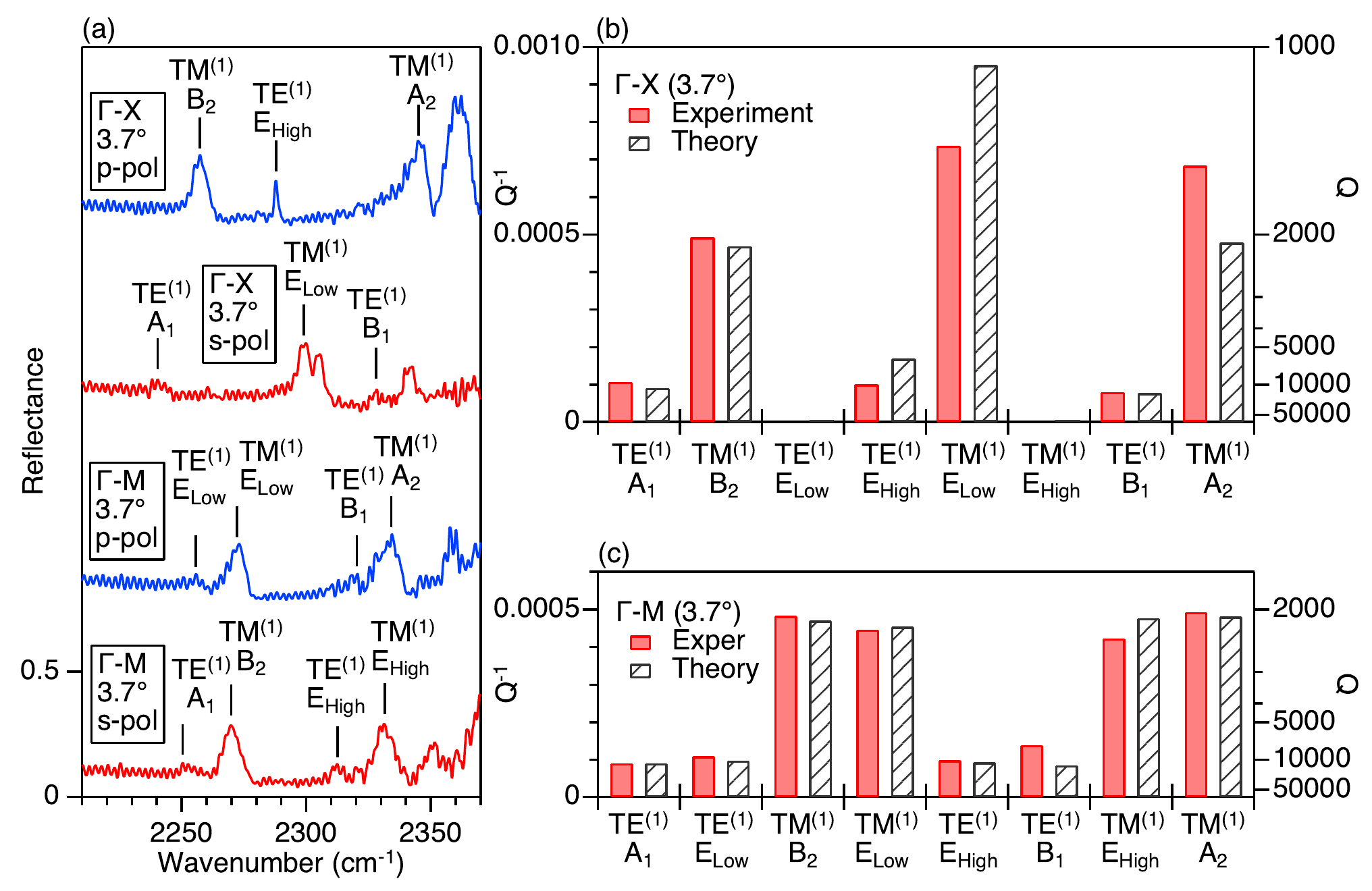}%
\caption{\label{fig_Qfactor}%
(a) Reflection spectra at $\theta = 3.7^{\circ}$ around the TE$^{(1)}$ and TM$^{(1)}$ frequency regions for different geometries and polarizations. The vertical lines indicate the mode center energies, which we determined via fitting. (b and c) Comparison of experimentally determined $Q^{-1}$ values (red filled bars) and theoretically obtained $Q^{-1}$ values (black hatched bars) for each mode depicted in (a) with (b) $\varGamma$-$X$ and (c) $\varGamma$-$M$ geometries. Note that the experimental $Q^{-1}$ values are proportional to the areas of the reflection peaks. %
}
\end{figure}

Figure~\ref{fig_Qfactor}(a) compares reflection spectra at $\theta = 3.7^{\circ}$ for different geometries and polarizations, where we focus on the TE$^{(1)}$ and TM$^{(1)}$ frequency regions. The vertical lines indicate the assigned peak energies, which we determined by fitting. The spectra indicate a strong intensity dependence on the mode index. %
(Reflection peaks of different modes have different intensities.) %
We can speculate that the peak intensity is related to the mode coupling strength, and thus the inverse of the $Q$ factor, i.e., the rate of energy leakage from the waveguide. The following consideration supports the validity of this speculation. 

A rigorous spectral analysis of 2D PC systems proves that the phase of electromagnetic waves transmitted through a PC slab is shifted by $\pi$ across the resonance frequency \cite{Ohtaka_PRB00}. Accordingly, the transmittance $|T|^2$ goes to zero at resonance, and the reflectance $|R|^2$ goes to unity, since $|R|^2=1-|T|^2$ for PC slabs made of non-absorbing materials. Thus, the area (spectrally-integrated intensity) of the reflection peaks roughly scales with the spectral width $\propto Q^{-1}$ independent of other factors, although the exact spectral shape is rather complicated due to interference between the resonant spectral component and nonresonant backgrounds. In reality, the ideally sharp spectral structures are masked due to the influence of PC inhomogeneities and distributed angles of incidence. Nevertheless, we are able to assume that the peak areas are inversely proportional to the mode $Q$ factors. 

Figure~\ref{fig_Qfactor}(b) shows experimentally determined and theoretically obtained $Q^{-1}$ values for all the TM$^{(1)}$ and TE$^{(1)}$ modes at $\theta = 3.7^{\circ}$. For the experimental determination we adopt a simple expression, namely, $Q^{-1}=(2/\pi)S/\omega_0$, where $S$ is the peak area, i.e., $\int |R|^2 d\omega$, and $\omega_0$ is the mode frequency. The coefficient ($2/\pi$) arises from the normalization factor of the Lorentzian function. The measured $Q^{-1}$ values reveal a fairly good agreement with the theoretical $Q^{-1}$ values, which we calculated using the finite element method, even without the use of any adjustable parameters. Hence, the reflection peak intensity provides a useful measure that is equivalent to the mode $Q$ factor, i.e., it constitutes key information for understanding lasing properties. 

Throughout this work we assume that refractive indices are real numbers and ignore the impact of absorption loss on the mode $Q$ factors. Hence, the $Q^{-1}$ values, which we analyzed theoretically and experimentally, are governed purely by diffraction (radiation) loss. The $Q^{-1}$ values are therefore equivalent with the rates of energy leakage from the waveguide to free space. To clarify the impact of material absorption on $Q$, we also calculate them using adequate complex refractive indices. The calculation results are shown by Table~\ref{tbl_QfactorsList} in Appendix. They indicate that the $Q$ factors calculated using complex indices ($Q_{\mathrm{total}}$) are significantly smaller than those calculated using real indices. The observed reduction in $Q_{\mathrm{total}}$ is due to free carrier absorption, which is not negligible for QCL devices in the mid-infrared region. Nevertheless, the present technique based on angle-resolved reflection makes it possible to determine the output coupling strengths, which are generally non-measurable if we use standard spectral techniques. %


\section{Conclusions}
In conclusion, we proposed angle-resolved reflection measurement with using a Fourier transform spectrometer as a useful technique with which to determine the in-plane dispersion relations of 2D PC slabs in the mid-infrared region. Polarization analysis enabled us to assign all the complex PC modes. The peak intensity evaluation enabled us to determine the mode $Q$ factors. Thus, the reflection measurement is useful for determining the optimal conditions for PCSEL structures. 

A potential problem inherent in the proposed technique is the limited angular resolution, and it might be rather difficult to distinguish fine splitting spectra at the $\Gamma$ point. However, the problem will be simply overcome by introducing a novel external-cavity QCL as an excitation source, since it generates a near-plane wave mid-IR beam. However, the frequency-tunable range is not so large. The adoption of our Fourier spectrometer-based technique for broad spectral analysis, and frequency-tunable QCLs for narrower spectral analysis constitute an ideal approach for fully characterizing PC slab samples that exhibit sharp resonance spectra. 
%


%
%

%

\begin{acknowledgments}
This work was supported by the Innovative Science and Technology Initiative for Security, Grant Number JPJ004596, ATLA, Japan. S.C. acknowledges support from the NIMS Graduate Assistantship program.
\end{acknowledgments}


\appendix*
\section{Analysis of electromagnetic field distributions}
This appendix shows the computed magnetic field distributions of our photonic crystal slab samples. They were calculated using the finite element method. 

\begin{figure}[b]
\centering
\includegraphics[width=8.5cm]{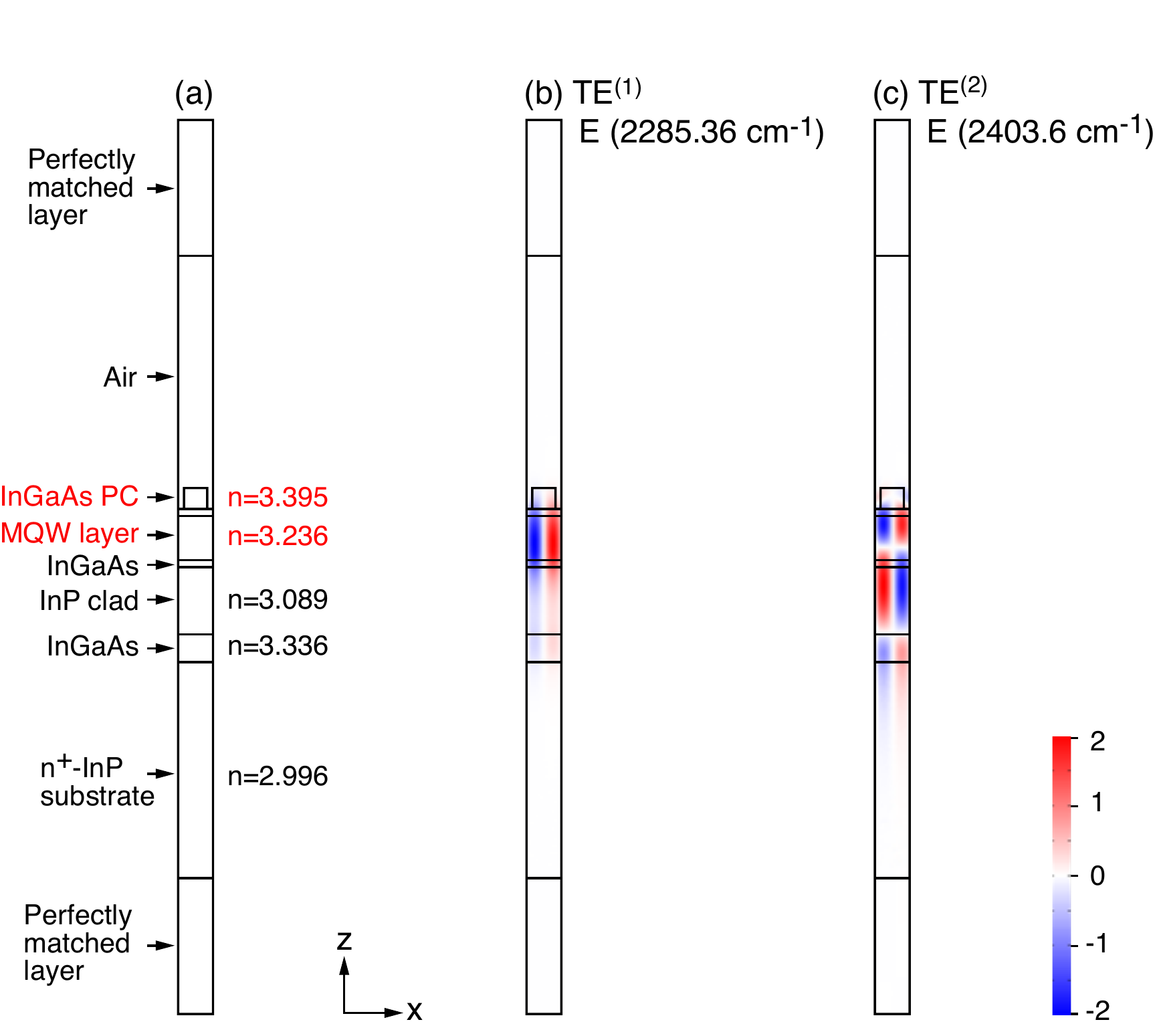}%
\caption{\label{fig_fieldDistributionVertical}%
Vertical field distributions. (a) Layer sequence used for our calculation. Note that the $z$-axis is normal to the slab surface. The refractive index $n$ of each layer is also indicated. The \textcolor{red}{red colored notations} highlight the waveguide core region. (b and c) The $z$ component of the magnetic field ($H_z$) for the lowest TE-like mode (TE$^{(1)}$) and the second lowest TE-like mode (TE$^{(2)}$), respectively. 
}
\end{figure}

\begin{figure}
\centering
\includegraphics[width=8.4cm]{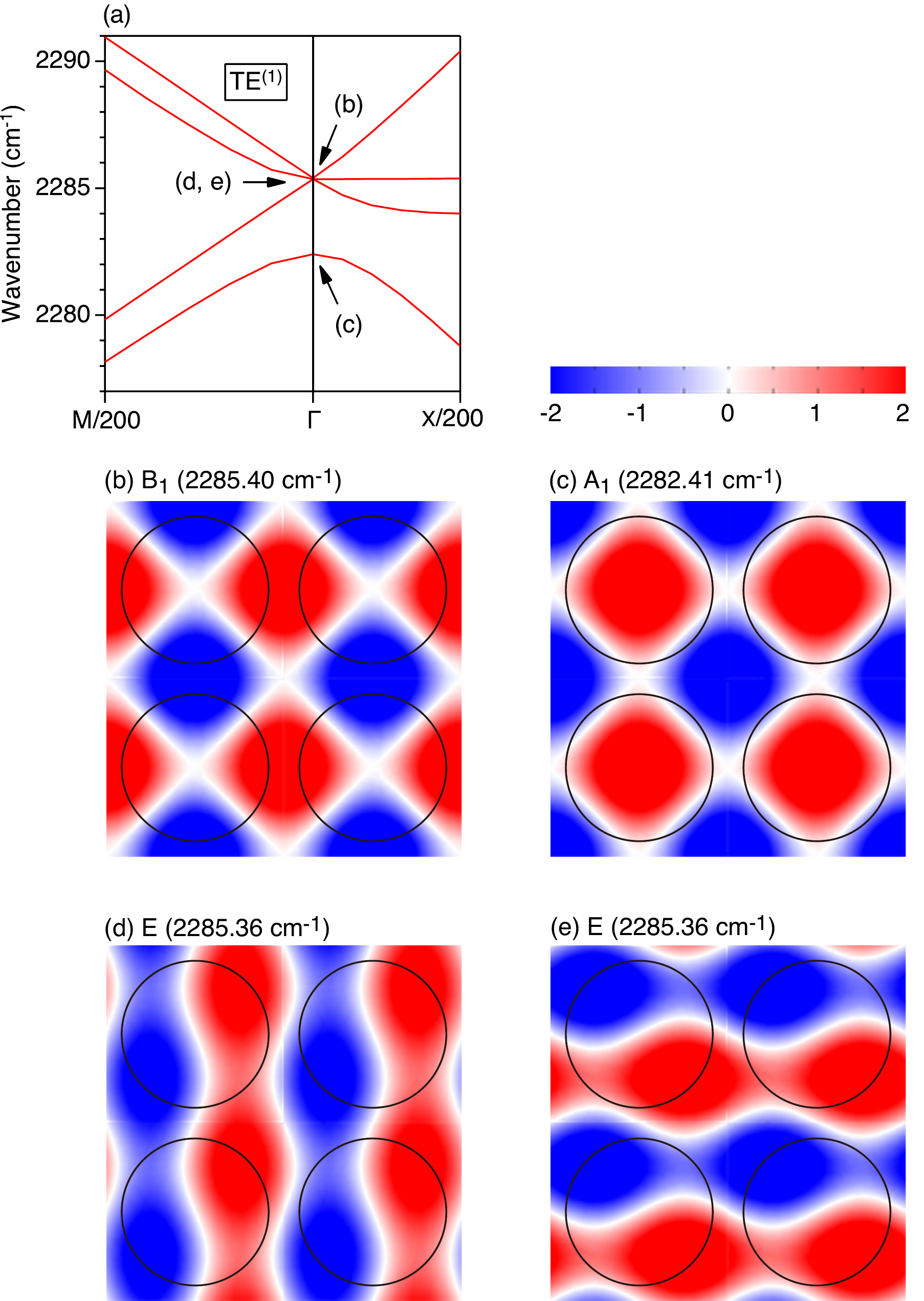}%
\caption{\label{fig_fieldDistributionLateral_TE}%
In-plane field distributions for the TE$^{(1)}$ modes at the $\varGamma$ point. (a) Dispersion relations in the $\varGamma$ point vicinity. (b-e) The $z$ component of the magnetic field ($H_z$) for modes depicted by the arrows in (a). The black lines indicate the boundary of circular pillars, which compose PC structures. All the in-plane distributions show their peculiar symmetries, which are characterized by (b) $B_1$, (c) $A_1$, and (d, e) $E$ for the $C_{4v}$ symmetry. %
Note that the dispersion relations in (a) suggest the degeneracy of $B_1$ and $E$ at the $\varGamma$ point and the formation of photonic Dirac cones \cite{Huang_NatComm2011,Sakoda_OEX12_2}, but in fact their eigenfrequencies are slightly different under the model condition. %
}
\end{figure}

Figure~\ref{fig_fieldDistributionVertical} is a comparison between the vertical field distributions of the lowest TE-like mode (TE$^{(1)}$) and the second lowest TE-like mode (TE$^{(2)}$). Figure~\ref{fig_fieldDistributionVertical}(a) shows the layer sequence of our sample and the refractive indices used for our calculation. Figure~\ref{fig_fieldDistributionVertical}(b) shows the field distribution of TE$^{(1)}$. It is strongly confined in the MQW layer along the $z$ axis, which is normal to the slab surface. 
Figure~\ref{fig_fieldDistributionVertical}(d) shows the field distribution of TE$^{(2)}$, which indicates a nodal distribution along $z$. See the dispersion relations of TE$^{(1)}$ and TE$^{(2)}$ in Fig.~\ref{fig_dispersion}(b) in the main body of this article, and the in-plane field distributions of TE$^{(1)}$ in Fig.~\ref{fig_fieldDistributionLateral_TE}(d). 

\begin{figure}
\centering
\includegraphics[width=8.4cm]{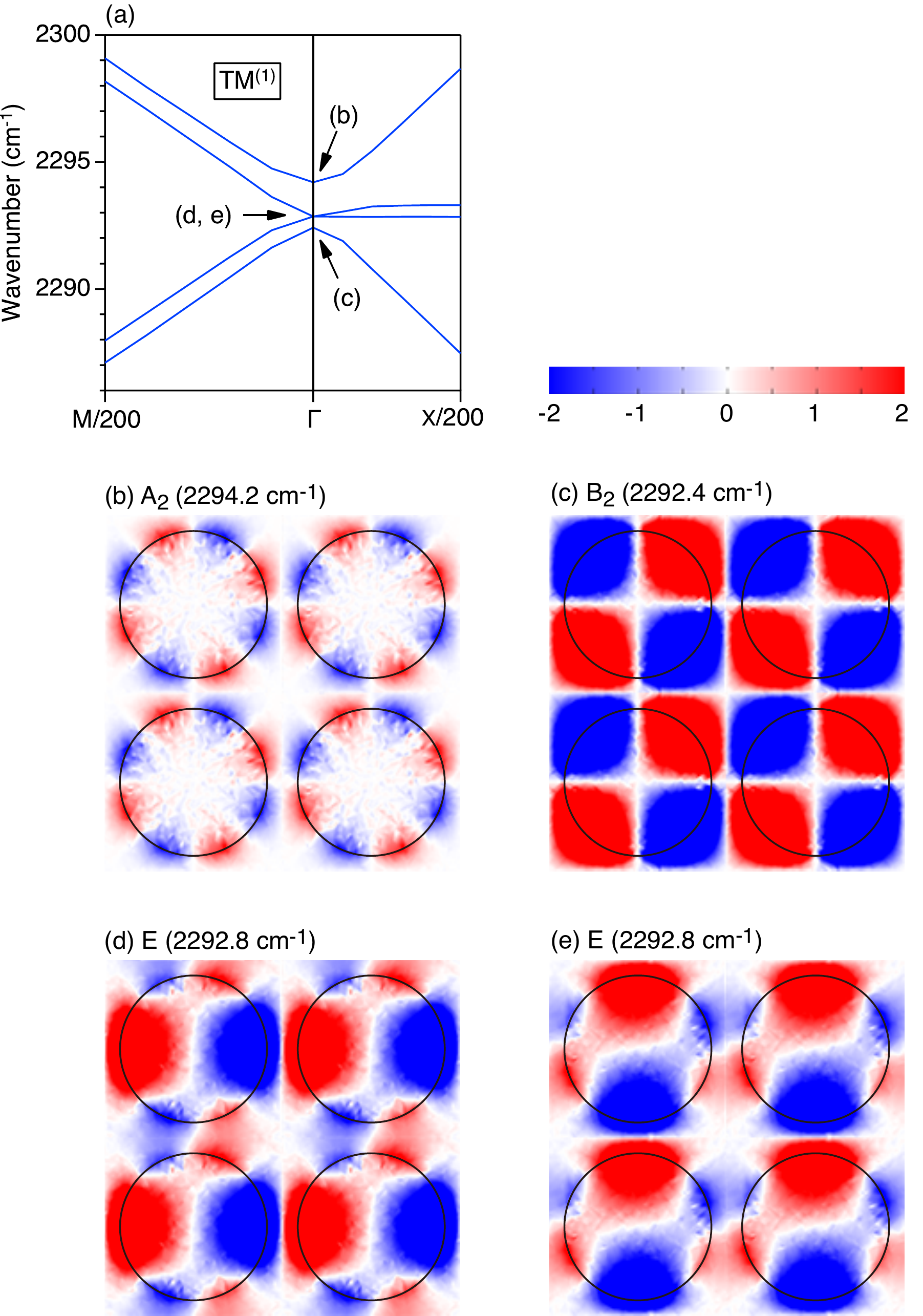}%
\caption{\label{fig_fieldDistributionLateral_TM}%
In-plane field distributions for the TM$^{(1)}$ modes at the $\varGamma$ point. (a) Dispersion relations in the $\varGamma$ point vicinity. (b-e) The $z$ component of the magnetic field ($H_z$) for modes depicted by the arrows in (a). The black lines indicate the boundary of circular pillars, which compose PC structures. Note that the relatively large fluctuation of image data is due to small $H_z$ values for the TM-like modes. Nevertheless, we can identify their spatial symmetries, which are characterized by (b) $A_2$, (c) $B_2$, and (d, e) $E$ for the $C_{4v}$ symmetry. %
}
\end{figure}

Figure~\ref{fig_fieldDistributionLateral_TE} shows the in-plane magnetic field distributions of the TE-like modes, and Figure~\ref{fig_fieldDistributionLateral_TM} shows those of the TM-like modes. All the in-plane distributions exhibit their peculiar symmetries, as can be seen in each panel. 


\begin{table}
\caption{\label{tbl_QfactorsList}
Comparison of the $Q$ factors calculated using real refractive indices and those calculated using complex refractive indices ($Q_{\mathrm{total}}$) for each TE$^{(1)}$ and TM$^{(1)}$ mode at (a) $k=X/50$ along the $\varGamma$-$X$ direction and (b) $k=M/50$ along the $\varGamma$-$M$ direction. Note that the $Q$ values calculated using real indices are used to plot Figs.~\ref{fig_Qfactor}(b) and \ref{fig_Qfactor}(c).}%
%
\begin{center}
\renewcommand{~}{\phantom{0}}
\begin{tabular}{lccc}
(a) $\varGamma$-$X$  ($k=X/50$) & & &\\ \hline\hline
 & Wavenumber  & $Q$        & $Q_{\mathrm{total}}$ \\ 
 & (cm$^{-1}$) & (real $n$)\footnotemark[1] & (complex $n$)\footnotemark[2] \\ \hline
TE$^{(1)}$ $A_1$ & 2262.4 & ~10968 & 2067 \\
TM$^{(1)}$ $B_2$ & 2270.9 & ~~2131 & 1131 \\
TE$^{(1)}$ $E$   & 2284.0 & 915694 & 2621 \\
TE$^{(1)}$ $E$   & 2285.5 & ~~5909 & 1788 \\
TM$^{(1)}$ $E$   & 2293.0 & ~~1050 & ~743 \\
TM$^{(1)}$ $E$   & 2293.4 &1064952 & 2489 \\
TE$^{(1)}$ $B_1$ & 2306.9 & ~12718 & 2211 \\
TM$^{(1)}$ $A_2$ & 2315.3 & ~~2087 & 1155 \\ \hline\hline%
\end{tabular}

\vspace{1\baselineskip}

\begin{tabular}{lccc}
(b) $\varGamma$-$M$  ($k=M/50$) & & &\\ \hline\hline
 & Wavenumber  & $Q$        & $Q_{\mathrm{total}}$ \\ 
 & (cm$^{-1}$) & (real $n$)\footnotemark[1] & (complex $n$)\footnotemark[2] \\ \hline
TE$^{(1)}$ $A_1$ & 2261.8 & 11026 & 2079 \\
TE$^{(1)}$ $E$   & 2263.3 & 10241 & 2012 \\
TM$^{(1)}$ $B_2$ & 2270.5 & ~2124 & 1147 \\
TM$^{(1)}$ $E$   & 2271.4 & ~2198 & 1131 \\
TE$^{(1)}$ $E$   & 2306.2 & 10689 & 2136 \\
TE$^{(1)}$ $B_1$ & 2307.7 & 11701 & 2159 \\
TM$^{(1)}$ $E$   & 2314.9 & ~2094 & 1174 \\
TM$^{(1)}$ $A_2$ & 2315.8 & ~2078 & 1138 \\ \hline\hline%
\end{tabular}
\end{center}
\footnotetext[1]{The list of the real $n$ values are shown in Fig.~\ref{fig_fieldDistributionVertical}(a).}
\footnotetext[2]{The imaginary parts of the complex $n$ values are estimated based on the doping levels of our QCL samples. }
\end{table}

\bibliography{midIRdisperion.bib}
\end{document}